\documentclass[aps,prl,twocolumn,superscriptaddress,preprintnumbers,nofootinbib,amsmath,amssymb,tighten]{revtex4} 
\usepackage{epsfig}
\usepackage{amsfonts}
\usepackage{amsmath}
\usepackage{amssymb}
\usepackage{dsfont}
\usepackage{hyperref}
\usepackage{color}
\usepackage{slashed}
\usepackage{multirow}

\newcommand{\al}{\alpha}

\newcommand{\g}{\gamma}

\newcommand{\simu}{\sigma^{\mu\nu}}

\newcommand{\sq}{^{2}}

\newcommand{\dslash}[1]{#1 \llap{/\kern-0.5pt}}
\newcommand{\Dslash}[1]{#1 \llap{/\kern+1.5pt}}
\newcommand{\DDslash}[1]{#1 \llap{/\kern+2.3pt}}
\newcommand{\dslashh}[1]{#1 \llap{/\kern+1pt}}

\newcommand{\bea}{\begin{eqnarray}}
\newcommand{\eea}{\end{eqnarray}}
\newcommand{\be}{\begin{equation}}
\newcommand{\ee}{\end{equation}}
\newcommand{\bma}{\begin{pmatrix}}
\newcommand{\ema}{\end{pmatrix}}
\newcommand{\nn}{\nonumber}

\begin{document}

\title{Is there room for CP violation in the top-Higgs sector?}
\author{ V. Cirigliano}
\affiliation{Theoretical Division, Los Alamos National Laboratory,
Los Alamos, NM 87545, USA}
\author{ W. Dekens}
\affiliation{Theoretical Division, Los Alamos National Laboratory, Los Alamos, NM 87545, USA}
\affiliation{New Mexico Consortium, Los Alamos Research Park, Los Alamos, NM 87544, USA}
\author{ J. de Vries}
\affiliation{Nikhef, Theory Group, Science Park 105, 1098 XG, Amsterdam, The Netherlands}
\author{ E. Mereghetti}
\affiliation{Theoretical Division, Los Alamos National Laboratory,
Los Alamos, NM 87545, USA}
\date{\today}

\preprint{NIKHEF 2016-011}
\preprint{LA-UR-16-21508}

\begin{abstract}

We  discuss direct and indirect  probes of chirality-flipping couplings of the top quark to Higgs and gauge bosons, considering both 
CP-conserving and CP-violating observables, in the framework  of the  Standard Model effective field theory. 
In our analysis we  include  current and prospective 
constraints from collider physics, precision electroweak tests, flavor physics, and electric dipole moments (EDMs).  
We find that low-energy indirect probes are very competitive, even after accounting for long-distance uncertainties. 
In particular,   EDMs put constraints on the electroweak CP-violating dipole moments  of the top that 
are two to three  orders of magnitude stronger than existing  limits. The new indirect constraint on the top EDM is given by
$|d_t| < 5 \cdot 10^{-20}$ e cm at $90\%$ C.L.

\end{abstract}
\maketitle

\textbf{Introduction:}  
The top quark  might  offer a  first gateway to  physics beyond the Standard Model (BSM), due
to  its large coupling to the Higgs and hence to the electroweak symmetry breaking  sector. 
In several scenarios, ranging from partial compositeness~\cite{Kaplan:1991dc}
to  supersymmetric  models with  light stops~\cite{Carena:2008vj}, 
enhanced  deviations from the SM  are expected  in the top sector which can be relevant for electroweak baryogenesis \cite{Carena:2008vj,Kobakhidze:2015xlz}.
Experiments at the Large Hadron Collider (LHC)  
offer a great opportunity to directly probe non-standard top quark couplings. 
On the other hand, these same  couplings also affect via quantum corrections 
processes that do not involve a top quark.  Such ``indirect probes'' give  
very valuable complementary information and 
in several cases constrain non-standard top couplings more strongly than direct searches. 

In this letter  we discuss  direct and indirect probes of  chirality-flipping 
top-Higgs couplings, including  both CP-conserving  (CPC) and CP-violating (CPV) interactions, the latter being 
of great interest in light of Sakharov's conditions for baryogenesis~\cite{Sakharov:1967dj}.
Despite the vast literature on 
top-gluon~\cite{Atwood:1992vj,Choudhury:2012np,Baumgart:2012ay,Biswal:2012dr,Bernreuther:2013aga,Hioki:2013hva,Aguilar-Saavedra:2014iga,Bramante:2014gda,Englert:2014oea,Rindani:2015vya,Gaitan:2015aia,Bernreuther:2015yna},
top-photon~\cite{CorderoCid:2007uc,Fael:2013ira,Bouzas:2012av,Bouzas:2013jha,Rontsch:2015una},  
top-$W$~\cite{Grzadkowski:2008mf,Drobnak:2010ej,GonzalezSprinberg:2011kx,Drobnak:2011aa,Cao:2015doa,Hioki:2015env,Schulze:2016qas},
top Yukawa~\cite{Brod:2013cka,Dolan:2014upa,Demartin:2014fia,Kobakhidze:2014gqa,Khatibi:2014bsa,Demartin:2015uha,Chen:2015rha,Buckley:2015vsa} couplings, 
and  global analyses~\cite{Kamenik:2011dk,Zhang:2012cd,deBlas:2015aea,Buckley:2015nca,Buckley:2015lku,Bylund:2016phk}, 
the impact of electric dipole moments (EDMs) has received comparatively little 
attention~\cite{CorderoCid:2007uc,Kamenik:2011dk,Brod:2013cka,Gorbahn:2014sha,Chien:2015xha}. 
The central new element of our work is the  systematic inclusion of  EDM constraints.  
Even after properly taking into account   the  hadronic and nuclear  uncertainties~\cite{Chien:2015xha},  
EDMs dominate the bounds on all  the CPV  top couplings. 
Our major finding is that  bounds on the  top EDM  (weak EDM) 
are improved by  three (two) orders of magnitude over the previous literature.
As part of our analysis, we also update indirect constraints from Higgs production and decay. 

We work in the linear SM Effective Field Theory (SM-EFT) 
framework~\cite{Buchmuller:1985jz,Grzadkowski:2010es,Jenkins:2013zja,Jenkins:2013wua, Alonso:2013hga}.
We assume that a gap exists between the scale of new physics $\Lambda$ and 
the electroweak scale $v=246$~GeV and keep only the  leading terms in $(v/\Lambda)^2$, 
corresponding to  dimension-six operators. 
We assume that at the high-scale  $\Lambda$ the largest non-standard  effects appear in the top sector, 
and hence set to zero all other couplings.
We then evolve the  non-standard top couplings to lower scales through renormalization group 
flow and heavy SM particle thresholds.  The evolution induces  operators 
that impact various high- and low-energy phenomena, 
thus leading to  constraints  on non-standard top-Higgs couplings  at the scale $\Lambda$. 

\textbf{Operator structure and  mixing pattern:}
In this letter we study   chirality-flipping couplings of the top quark to Higgs and gauge bosons.
At dimension six,  five structures arise: 
non-standard Yukawa and  Gluon, Electric, and Weak dipoles.
The hierarchical flavor structure considered here 
naturally arises~\cite{Cirigliano:2016nyn}  in models obeying 
minimal flavor violation (MFV)~\cite{D'Ambrosio:2002ex}. 
The starting effective  Lagrangian  encoding  new physics  at the high scale $\Lambda \gg v$ 
in the quark mass basis is  
\begin{equation}
{\cal L}_{\rm eff}^{\rm BSM}   =
\sum_{\alpha \in \{ Y, g, \gamma, Wt,Wb\} }  \ C_\alpha  \, O_\alpha + {\rm h.c.} \qquad  \quad
\label{eq:Leff}
 \end{equation}
with complex couplings  $C_\alpha = c_\alpha + i \,   \tilde{c}_\alpha$  and
\begin{subequations}
\label{eq:operators}
\bea
O_Y  &=&   -  m_t  \bar{t}_L  t_R  \left( v h  + \frac{3}{2}  h^2 + \frac{1}{2}  \frac{h^3}{v}  \right)   
\\
O_\gamma &=&  - \frac{e Q_t}{2}  m_t  \, \bar{t}_L  \sigma_{\mu \nu} \left( F^{\mu \nu} - t_W  Z^{\mu \nu} \right) t_R \, \left(1 + \frac{h}{v}\right)   
\qquad 
\\
O_g &=&  - \frac{g_s}{2}  m_t  \, \bar{t}_L  \sigma_{\mu \nu} G^{\mu \nu} t_R \, \left(1 + \frac{h}{v}\right)   
\\
O_{Wt} \!\!   &=& -g m_t \bigg[  \frac{1}{\sqrt{2}}  \bar{b}_L'   \simu  t_R W_{\mu\nu}^- 
\nn \\
&+& \!  \! \bar t_L\simu t_R \bigg(\frac{1}{2c_W} Z_{\mu\nu}+i g W_\mu^-W_\nu^+\bigg)\bigg]\bigg(1+\frac{h}{v}\bigg) \qquad
\\
O_{Wb} \!\! &=& - g m_b   \bigg[\frac{1}{\sqrt{2}} \bar t_L'  \simu b_R  W_{\mu\nu}^+ 
\nn \\
&-& \! \! \bar b_L\simu b_R \bigg(\frac{1}{2c_W} Z_{\mu\nu}+i g W_\mu^-W_\nu^+\bigg)\bigg]\bigg(1+\frac{h}{v}\bigg), \qquad
\eea
\end{subequations}
where $Q_t = 2/3$,  $t_W = \tan \theta_W$,  $c_W = \cos \theta_W$,  
$b' = V_{tb} b + V_{ts} s + V_{td} d$, $t' = V_{tb}^* t + V_{cb}^* c + V_{ub}^* u$, and $h$ denotes the physical scalar boson.
Our operators retain the full constraints of gauge invariance as they are 
linear combinations of the explicitly $SU(2) \times U(1)$-invariant operators 
of Refs.~\cite{Buchmuller:1985jz,Grzadkowski:2010es}, expressed in the unitary gauge.  
The correspondence to the standard basis is provided in Table~\ref{tab:op-invariant}.
The couplings  $C_\alpha$ have mass dimension $[-2]$ and are related to  properties of the 
top quark, such as electric and  magnetic dipole moments 
($d_t = (e m_t Q_t)  \tilde{c}_\gamma$ and  $\mu_t = (e m_t Q_t)  {c}_\gamma$).

\begin{table}
\scriptsize
$\begin{array}{|c|c|c|}
\hline 
\multicolumn{2}{|c|}{\rm Operator}   &  {\rm Coupling}   \\
\hline
\hline
- \sqrt{2}  \varphi^\dagger \varphi  \  \bar{q}_L   \,  Y_u'  \, u_R  \, \tilde  \varphi  &   O_Y  & 
y_t  C_Y  =   [Y_u']_{33} 
\\   \hline 
- \frac{g_s}{\sqrt{2}}  \  \bar{q}_L  \sigma \cdot G   \, \Gamma_g^u  \, u_R \, \tilde  \varphi &      O_g & 
y_t C_g =  [\Gamma_g^u]_{33}
\\  \hline \hline
- \frac{g'}{\sqrt{2}}  \  \bar{q}_L  \sigma \cdot B   \, \Gamma_B^u  \, u_R \, \tilde  \varphi  &    O_{\gamma,Wt}    & 
y_t  Q_t C_\gamma = -  [ \Gamma_B^u + \Gamma_W^u ]_{33} 
\\  
- \frac{g}{\sqrt{2}}  \  \bar{q}_L  \sigma \cdot W^a \tau^a   \, \Gamma_W^u  \, u_R \, \tilde  \varphi  &      &  
y_t C_{Wt} = [\Gamma_W^u]_{33} 
\\ \hline  \hline
- \frac{g}{\sqrt{2}}  \  \bar{q}_L  \sigma \cdot W^a \tau^a   \, \Gamma_W^d  \, d_R \,   \varphi  &    O_{Wb} & 
y_b C_{Wb} = [\Gamma_W^d]_{33} 
\\ \hline
\end{array}$
\caption{High-scale operators in $SU(2)\times U(1)$ invariant form~\cite{Buchmuller:1985jz,Grzadkowski:2010es} 
(left column) and mapping to the operators and couplings used in this letter (center and right column).
$q_L$ represents the L-handed quark doublet, $\varphi$ is the Higgs doublet, and $\tilde \varphi = i \sigma_2 \varphi^*$.   
$g_s, g, g'$ denote the $SU(3)$, $SU(2)$,  and $U(1)$ gauge couplings, 
$y_{t,b} = m_{t,b}/v$,  
and  $\sigma \cdot X = \sigma_{\mu \nu} X^{\mu \nu}$.
The couplings $C_{\alpha}$ are related to the  $33$ components of the 
matrices $Y_u'$ and  $\Gamma_{g,B,W}^{u,d}$  in the quark mass basis. 
} \label{tab:op-invariant}
\end{table}

To   constrain   $c_\alpha$ and $\tilde{c}_\alpha$  we use direct and indirect probes. 
Direct probes involve top quark production 
(single top,  $t \bar t$,  and $t \bar t h$) and  decay  ($W$-helicity fractions, lepton angular distributions) at colliders. 
We include CPV effects in the angular distributions of the decay products of a single top  \cite{Aad:2015yem}, 
while we neglect CPV observables in 
$t \bar t$ and $t \bar t h$ production/decay~\cite{Bernreuther:1992be,Brandenburg:1992be,Bernreuther:1993hq,Choi:1997ie,Sjolin:2003ah,Antipin:2008zx,Gupta:2009wu,Gupta:2009eq,Hayreter:2015ryk}  as these are not yet competitive.
Indirect probes involve top quarks in quantum loops, 
affecting both high-energy  (Higgs production and decay, precision electroweak tests) and  low-energy observables
($b \to s \gamma$  and EDMs).

Indirect constraints rely on operator-mixing via renormalization group (RG) flow and  on threshold corrections 
arising from integrating out heavy SM particles ($t,h,W,Z$).
In Table~\ref{tab:extended} we summarize the operators 
that are generated  from Eq.~(\ref{eq:Leff}) 
to leading order in the strong, electroweak, and Yukawa couplings.
These  include the  light quark  electromagnetic and gluonic dipoles (flavor diagonal  
and off-diagonal entries relevant to $b \to s \gamma$),  the Weinberg three-gluon operator,  
and operators involving Higgs and  gauge bosons.
To a good approximation, these operators close under RG evolution, 
and in particular the top dipoles  mix into  and from  chirality-conserving top-Higgs-gauge  
couplings at three-loops~\cite{Grzadkowski:2010es,Jenkins:2013zja,Jenkins:2013wua, Alonso:2013hga}.

There are several  paths to connect the   high-scale Wilson coefficients in (\ref{eq:Leff}) to the 
operators in Table~\ref{tab:extended} and low-energy observables. 
These paths are  determined  by the RG equations 
\bea
\frac{d C_i}{d\ln \mu} = \sum_j \g_{j\rightarrow i} \, C_j ~, 
\label{eq:RG1}
\eea
and possibly threshold corrections. 
In Table~\ref{tab:observables} we provide a synopsis  
of the  induced  low-scale couplings (left column) and  
the  observables they contribute to (right column). 
Several of these  paths have already  been analyzed in the literature.  
Here we briefly recall the dominant paths for each operator, 
paying special attention to a novel  two-step path that connects the 
top EDM and W-EDMs  ($\tilde{c}_\gamma$ and  $\tilde{c}_{Wt}$) 
to low energy. A detailed  analysis is presented in Ref.~\cite{Cirigliano:2016nyn}.

\begin{figure} 
\includegraphics[width=.9\linewidth]{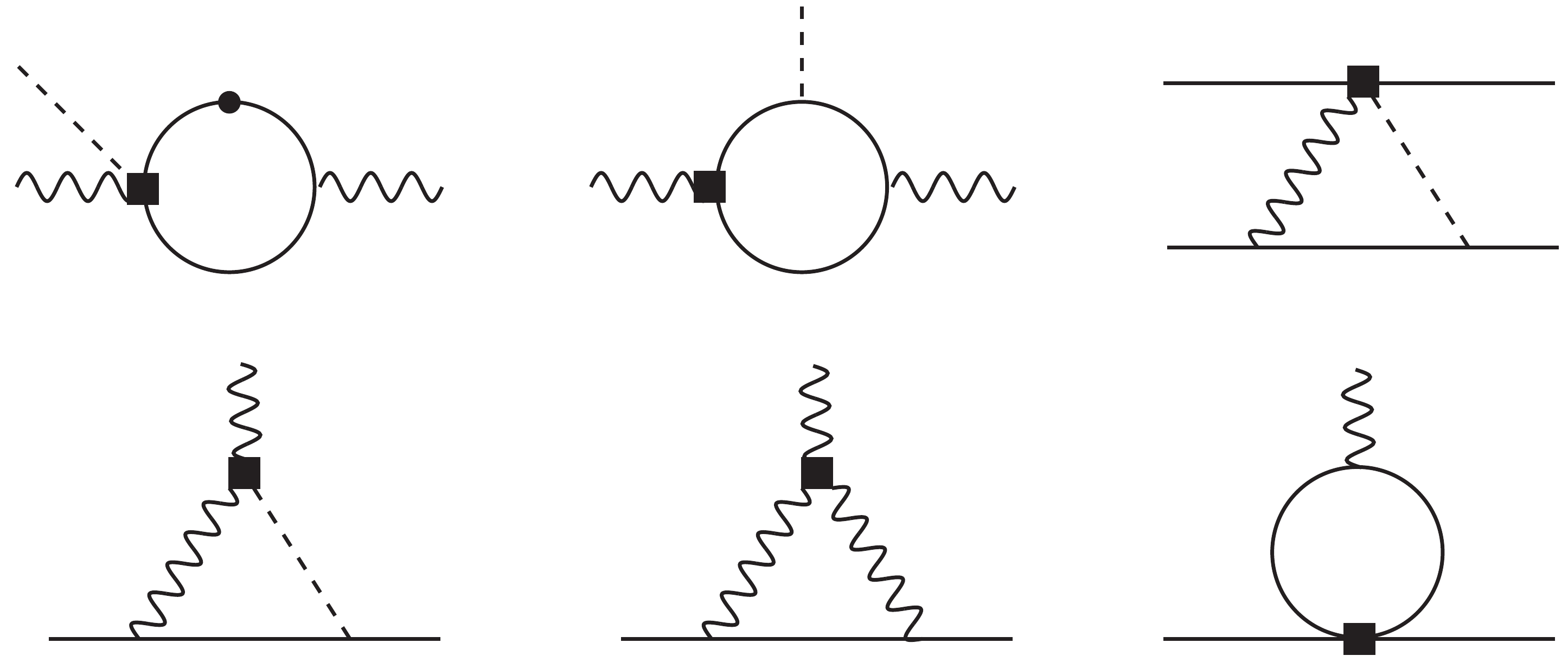}
\caption{
Representative diagrams contributing to the mixing of $C_\gamma$ into 
$C_{\varphi \tilde W, \varphi \tilde B, \varphi \tilde W B,quqd,lequ}$ (top panel), and 
the mixing of the latter into  light fermion  electroweak dipoles (bottom panel). 
The  square  (circle) represents an operator  (quark mass) insertion.
Solid, wavy, and dotted lines represent fermions,  electroweak gauge bosons, and the Higgs, respectively.  
}
\label{fig:diagrams}
\end{figure}

There are three paths that constrain the top electromagnetic dipole coupling  $C_\gamma$ 
through indirect  measurements. 
First,  $C_\gamma$  induces down-type EDMs  ($C_\gamma \to C_\gamma^{(d,s)}$)  via a flavor-changing  $W$ loop,  
suppressed by the CKM factor $|V_{td,ts}|^2$. 
Similar one-loop diagrams induce $b \to s \gamma$ dipole operators \cite{Aebischer:2015fzz,Hewett:1993em}. 
Next,  at one loop $C_\gamma$ induces the top gluonic dipole $C_g$,  which 
in turn  at the top threshold generates  the three-gluon Weinberg coupling $C_{\tilde G}$.  
Finally, there is a new two-step path:  first $C_\gamma$ induces 
the anomalous couplings of the Higgs to electroweak bosons, as well as anomalous couplings of the top quark to light fermions, namely 
$C_{\varphi W, \varphi B, \varphi W B}$, $C_{\varphi \tilde W, \varphi \tilde B, \varphi \tilde W B}$, and $C_{lequ,quqd}$ (see top diagrams in Fig.~\ref{fig:diagrams}).
These couplings in turn  mix  at one loop   (see bottom diagrams in Fig.~\ref{fig:diagrams})  into the electromagnetic dipoles 
$C_\gamma^{(f)}$  ($f= e,\,u,\,d,\,s$) \cite{McKeen:2012av,Dekens:2013zca}. 
Focusing on the electron, the relevant anomalous dimensions are 
\begin{equation}
\gamma_{\tilde{c}_\gamma  \to   C_{\{\varphi \tilde B , \varphi \tilde W B , lequ\} } }=
\frac{N_c  Q_t }{16 \pi^2}y_t\sq   \, \{1- 4 Q_t     , 1, \frac{3g^{\prime 2} y_e}{2N_c y_t}\}~,
\end{equation} 
\be
\gamma_{ \{  C_{\varphi \tilde B} , C_{\varphi \tilde W B} , C_{lequ}^{(3)}\} \to  \tilde{c}_\gamma^{(e)}}  \ = \  
-\frac{\alpha}{\pi \, Q_e \, s_W^2}  \ \times
\nonumber 
\ee
\be
\left\{ 
-3 t_W^2  ,        \
\frac{3}{2}(1+ t_W^2)    , 4Q_tN_c \frac{1}{g\sq}\frac{y_t}{ y_e }
\right\} ~.
\ee
This new ``two-step'' path leads to light fermion EDMs.   For the electron, 
the approximate solution of (\ref{eq:RG1}) reads
\be
\frac{\tilde{c}_\gamma^{(e)}  }{\tilde{c}_\gamma} \simeq   \frac{3 N_c Q_t \alpha }{64 \pi^3 s_W^2}  \frac{m_t^2}{v^2}  
\, \left[ 1 + (12 Q_t - 1)  t_W^2 \right]  \left( \log \frac{\Lambda}{m_t} \right)^2,
\ee
implying $\tilde{c}_\gamma^{(e)}/\tilde{c}_\gamma \sim 4 \times 10^{-4}$ for $\Lambda = 1$~TeV
and thus $|d_e| \sim | v^2 \tilde{c}_\gamma |\ \cdot 6 \cdot 10^{-26}  e \, {\rm cm}$. 
While  this simple estimate already shows the power of this new path
($|d_e| < 8.7 \cdot 10^{-29} e\, {\rm cm}$~\cite{Baron:2013eja}),
in our analysis we employ the full numerical solution of (\ref{eq:RG1}).

The  weak dipole $C_{Wt}$   has a  mixing pattern similar to $C_\gamma$. 
The strongest  constraints arise again from the two-step path: 
$C_{Wt}$
$\to $ 
$C_{\varphi \tilde W, \varphi \tilde B, \varphi \tilde W B,lequ,quqd}$  $\to$ 
$C_\gamma^{(f)}$  ($f= e,\,u,\,d,\,s$). 
For $C_{Wb}$  this path is  suppressed by the bottom Yukawa. So the main contribution of $C_{Wb}$
to EDMs arises from  mixing with the  $b$ chromo-EDM, which induces $O_{\tilde G}$~\cite{BraatenPRL,Boyd:1990bx}  at the $m_b$ threshold.

The gluonic dipole coupling $C_g$   mixes at one loop with the top electromagnetic dipole $C_\gamma$,  
the non-standard Yukawa  $C_Y$,  and non-standard Higgs-gluon 
couplings $C_{\varphi G,\varphi \tilde G}$ \cite{Degrassi:2005zd,Dekens:2013zca,Degrande:2012gr,Alonso:2013hga,Chien:2015xha}.  
Moreover, $C_g$ generates the light chromo-EDMs through the two-step mechanism, $C_{g}$
$\to $ 
$C_{quqd}^{(1),(8)}\to C_g^{(q)} $, and induces $O_{\tilde G}$ at the top threshold. 

Finally, the non-standard top Yukawa coupling $C_Y$ has no anomalous mixing but 
it contributes to  all  the couplings of 
the extended effective Lagrangian at lower scale  through finite threshold corrections 
from one-loop and two-loop Barr-Zee diagrams \cite{Barr:1990vd,Gunion:1990iv,Abe:2013qla,Jung:2013hka,Dekens:2014jka,Weinberg:1989dx,Dicus:1989va}. 

\begin{table}
\small
$\begin{array}{|c|}
\hline
\hline
O_{\varphi G} = g_s^2 \varphi^\dagger \varphi   G_{\mu \nu} G^{\mu \nu}  
\qquad 
O_{\varphi \tilde G}= g_s^2\varphi^\dagger \varphi    G_{\mu \nu} \tilde G^{\mu \nu}  
\\ \hline
O_{\varphi W} = g^2 \varphi^\dagger \varphi   W^i_{\mu \nu} W^{i \mu \nu} 
\qquad
{O}_{\varphi \tilde W} = g^2 \varphi^\dagger \varphi  \, \tilde W^i_{\mu \nu} W^{i \mu \nu} 
\\ \hline
O_{\varphi B} = g'^2 \varphi^\dagger \varphi   B_{\mu \nu} B^{\mu \nu} 
\qquad 
{O}_{\varphi \tilde B}= g'^2\varphi^\dagger \varphi     \,  \tilde B_{\mu \nu} B^{\mu \nu} 
\\ \hline
O_{\varphi W B} = g g'\varphi^\dagger \tau^i  \varphi   W^i_{\mu \nu}  B^ {\mu \nu} 
\qquad 
{O}_{\varphi \tilde W B}= g g' \varphi^\dagger \tau^i  \varphi     \,  \tilde W^i_{\mu \nu}  B^ {\mu \nu} 
\\ \hline 
O_{lequ}^{(3)}=(\bar l_L^I\simu e_R) \epsilon_{IJ}(\bar q^J_L\sigma_{\mu\nu}u_R)\\\hline
O_{quqd}^{(1)}=(\bar q_L^I u_R) \epsilon_{IJ} (\bar q_L^J d_R),\quad O^{(8)}_{quqd}=(\bar q_L^I\, t^a\, u_R) \epsilon_{IJ} (\bar q_L^J\, t^a\, d_R)
\\ \hline
O_{\tilde G} = (1/6)  g_s  f_{abc} \epsilon^{\mu \nu \alpha \beta}  G^{a}_{\alpha \beta}  G^{b}_{\mu \rho} G^{c \  \rho}_{\nu} 
\\ \hline
O_g^{(q)} =   O_g \vert_{t \to q} \qquad q =u,d,s 
\\ \hline
O_g^{(bs)} =   - (g_s/2)  \, m_b  \, \bar{s}_L \sigma_{\mu \nu} G^{\mu \nu}  b_R   
\\ \hline 
O_\gamma^{(f)} =   O_\gamma \vert_{t \to f}   \qquad f =  e, u,d,s  
\\ \hline
O_\gamma^{(bs)} =   - (e/2 \, ) m_b  \, \bar{s}_L \sigma_{\mu \nu} F^{\mu \nu}  b_R   
\\\hline
\hline
\end{array}$
\caption{
Dimension-six operators induced by the  
top-Higgs interactions in Eq.~(\ref{eq:Leff}) via RG flow and threshold corrections.
We write $\tilde{X}_{\mu \nu} \equiv \epsilon_{\mu \nu \alpha \beta}  X^{\alpha \beta}$.
} \label{tab:extended}
\end{table}

\begin{table}[t]
\small
$\begin{array}{|c|c|}
\hline 
{\rm Coupling}  &    {\rm  Observables}   \\
\hline
\hline
C_g &  \sigma (t \bar t);  \ \sigma (t \bar t h)
\\ \hline
C_{Wt} &  \sigma (t);   \ t \to W \, b
\\ \hline
C_{Wb} &  \sigma (t);   \ t \to W \, b ; \  Z \to b \bar{b}
\\ \hline
C_{Y} &  \sigma (t \bar t h)  
\\ \hline 
\hline
C_{\varphi W, \varphi B,\varphi WB}    \leftarrow  \  C_\gamma,  C_{Wt,Wb} , C_Y   &   h \to \gamma \gamma; ~ S 
\\ \hline 
C_{\varphi  \tilde W, \varphi  \tilde B,\varphi \tilde W B}    \leftarrow  \  C_\gamma,  C_{Wt,Wb} , C_Y   &    h \to \gamma \gamma 
\\ \hline 
C_{\varphi G, \varphi \tilde G}       \leftarrow  \  C_g,   C_Y   &   h  \leftrightarrow g g 
\\ \hline 
\hline
C_{\tilde G}    \leftarrow  \  C_g,   C_Y   &   {\rm EDMs}
\\ \hline 
C_{g}^{(q)} \leftarrow  \  C_{\al}, C_{\varphi G,  \varphi \tilde G}  , C_{quqd}^{(1,8)}    &  {\rm EDMs}; \  b \to s \gamma
\\ \hline 
C_{\gamma}^{(f)} \leftarrow  \  C_{\al\neq g}, C_{lequ}^{(3)}, C_{quqd}^{(1),(8)} ,  \qquad    &  {\rm EDMs}; \  b \to s \gamma
\\
\qquad \quad C_{\varphi W, \varphi B, \varphi W B}, \,  C_{\varphi \tilde W, \varphi \tilde B, \varphi \tilde W B}    & 
\\ \hline 
\end{array}$
\caption{Left column:  effective couplings in the extended basis. 
The first four entries are in the original basis of Eq.~(\ref{eq:Leff}). 
The  remaining  entries are induced via  RG flow, as indicated by the arrows. 
Right column:  observables to which couplings in the extended basis contribute.  
} \label{tab:observables}
\end{table}

\begin{figure*}[t!]
\includegraphics[width=0.47\linewidth]{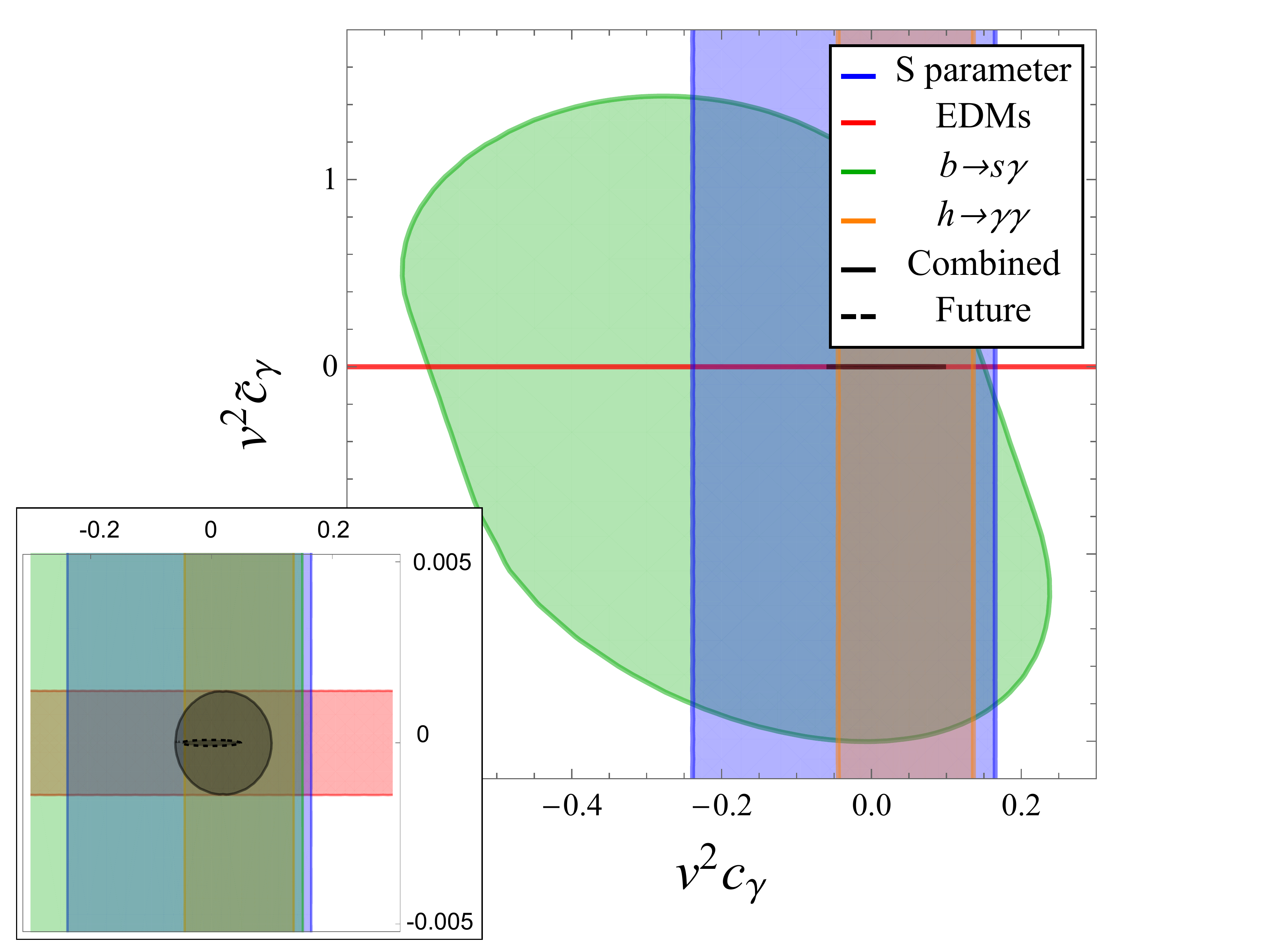}
\includegraphics[width=0.47\linewidth]{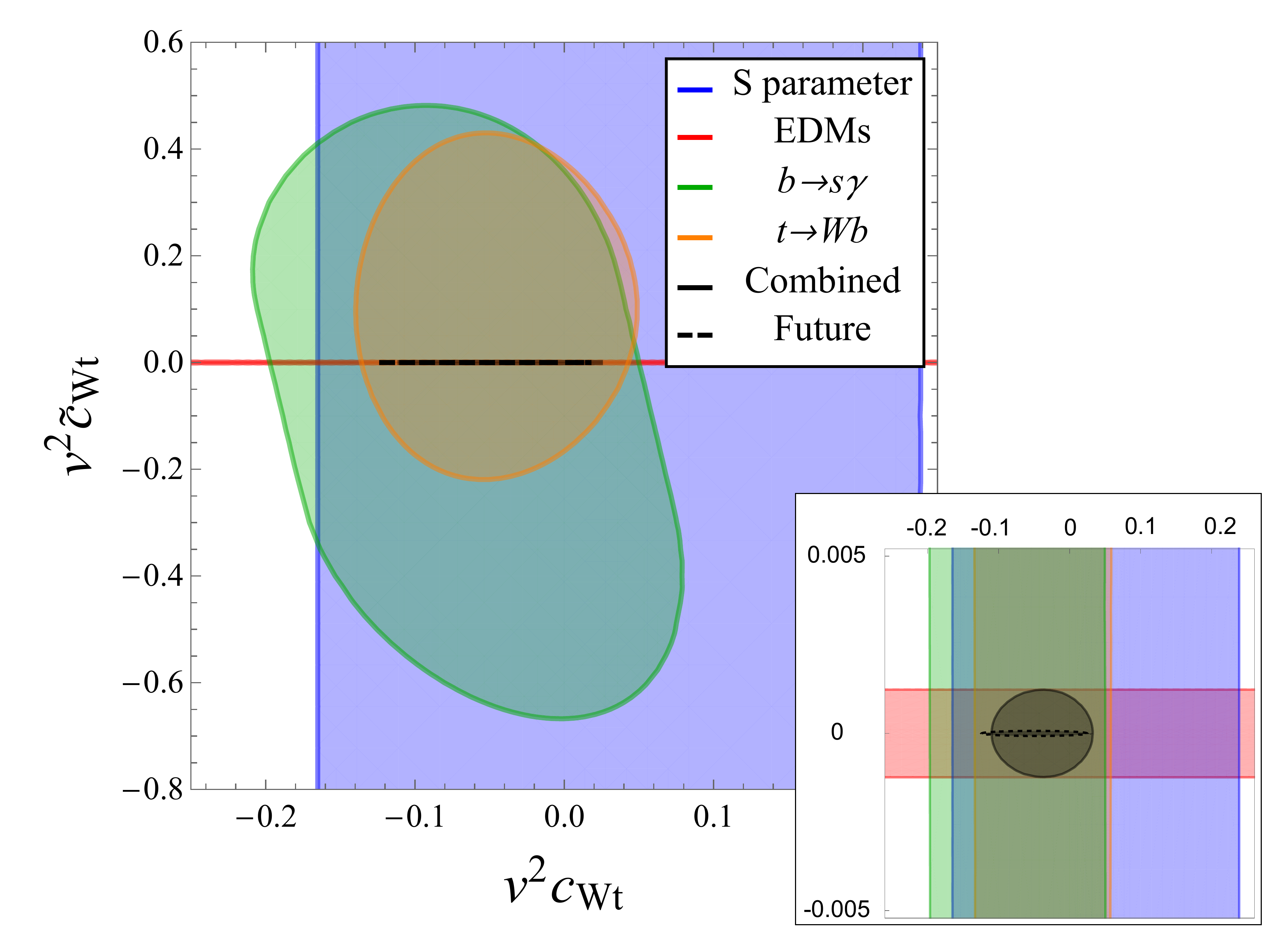}
\caption{90\% CL  allowed regions in the   $v^2 c_{\gamma} - v^2  \tilde{c}_{\gamma}$   (left panel) 
and $v^2 c_{Wt} - v^2  \tilde{c}_{Wt}$  planes (right panel), with couplings evaluated at   $\Lambda = 1$~TeV.
In both cases, the inset  zooms into the current combined allowed region and shows projected future sensitivities. 
Future EDM searches will probe  $v^2 \tilde{c}_{\gamma} \sim  8 \cdot 10^{-5}$ and $v^2 \tilde{c}_{Wt} \sim 7 \cdot 10^{-5}$.
\label{fig:CgammaCWt}}
\end{figure*}

\textbf{Current and prospective bounds:}
As  becomes clear from Table~\ref{tab:observables},  
the high-scale top-Higgs couplings can be constrained by various CP-even and CP-odd  observables.   
A detailed description of the experimental  and theoretical input, the chi-squared function, and the treatment of theoretical 
uncertainties are presented in Ref.~\cite{Cirigliano:2016nyn}.
Here we highlight  the main features of our analysis:
(i) For each observable, we include only contributions linear  in the 
new physics couplings $C_\alpha$,
neglecting higher-order terms in the SM-EFT expansion. 
We express all bounds in terms of $C_\alpha (\Lambda = 1 \,{\rm TeV})$. 
(ii) For low-energy probes ($b \to s \gamma$ 
 \cite{Altmannshofer:2012az,Altmannshofer:2011gn,Lunghi:2006hc,Benzke:2010tq} and EDMs~\cite{Chien:2015xha,Engel:2013lsa,Pospelov_review})  
we treat the significant hadronic and nuclear  theoretical uncertainties according to the ``range-fit'' method~\cite{Charles:2004jd}, 
in which  the total chi-squared is minimized with respect to the matrix elements (varied in their allowed theoretical range). 
This procedure allows for cancellations between different contributions to a given observable and 
thus gives the most conservative bounds on BSM couplings~\cite{Chien:2015xha}. 
(iii)  We use experimental input on top processes at Tevatron~\cite{Aaltonen:2012rz,Aaltonen:2013wca}
and the LHC~\cite{Aad:2014kva,Chatrchyan:2013faa,Aad:2014fwa,Khachatryan:2014iya,Aad:2012ky,Aad:2015yem,Khachatryan:2014vma,ATLAS-CONF-2015-079,CMS-PAS-TOP-15-004}; 
on Higgs production/decay signal strengths~\cite{Aad:2015gba,Khachatryan:2014jba}; 
on $Z \to b \bar b$ and $S$~\cite{ALEPH:2005ab,Agashe:2014kda}; 
on $ b \to s \gamma$~\cite{Amhis:2014hma,Agashe:2014kda}; 
and on  neutron,  $^{199}$Hg,  $^{129}$Xe,  $^{225}$Ra, 
and electron EDMs~\cite{Baker:2006ts,Afach:2015sja,PhysRevLett.86.22,Parker:2015yka,Graner:2016ses,Baron:2013eja}.

\begin{figure*}[t]
\includegraphics[width=0.3\linewidth]{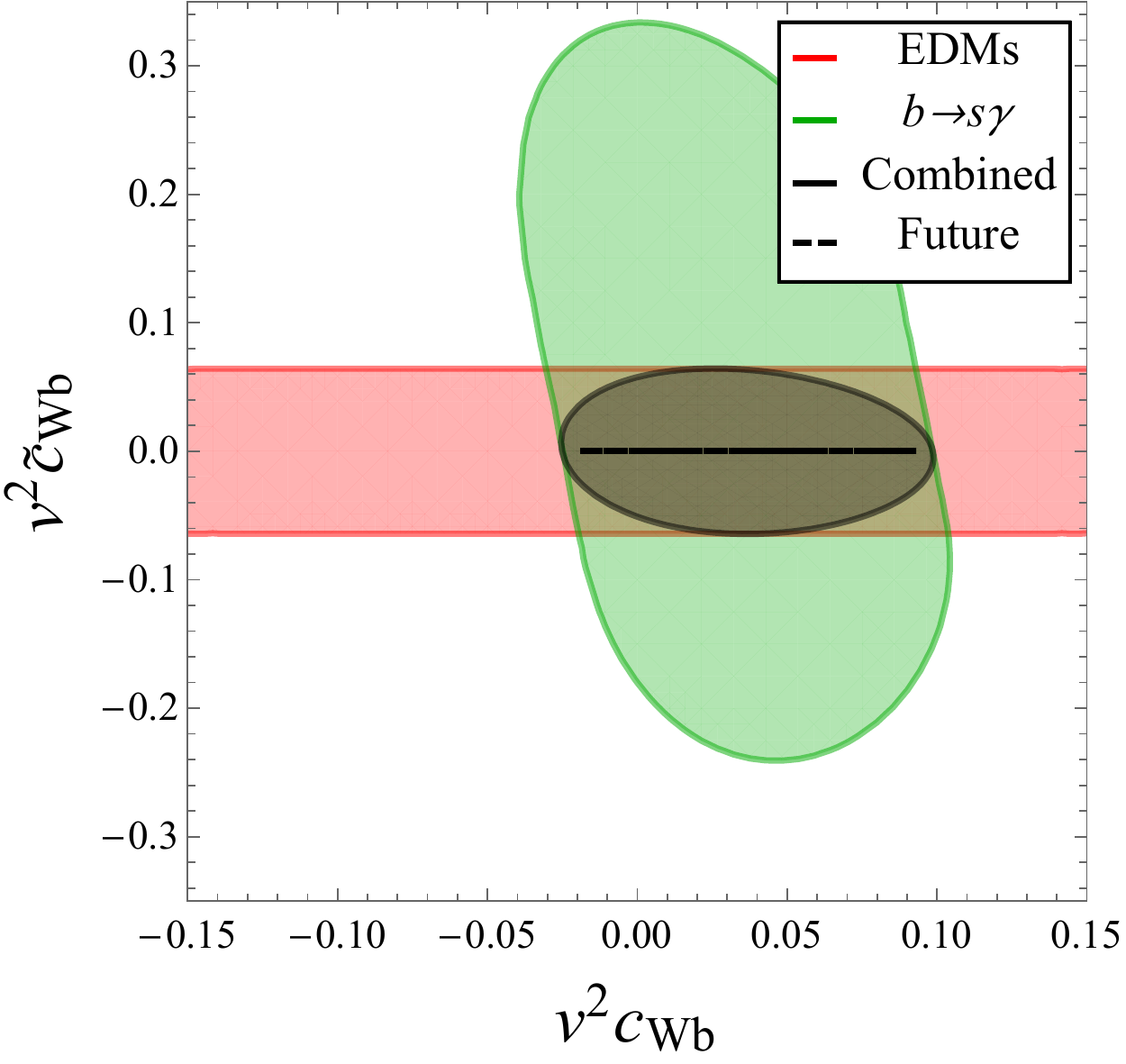}
\hspace{0.03\linewidth}
\includegraphics[width=0.3\linewidth]{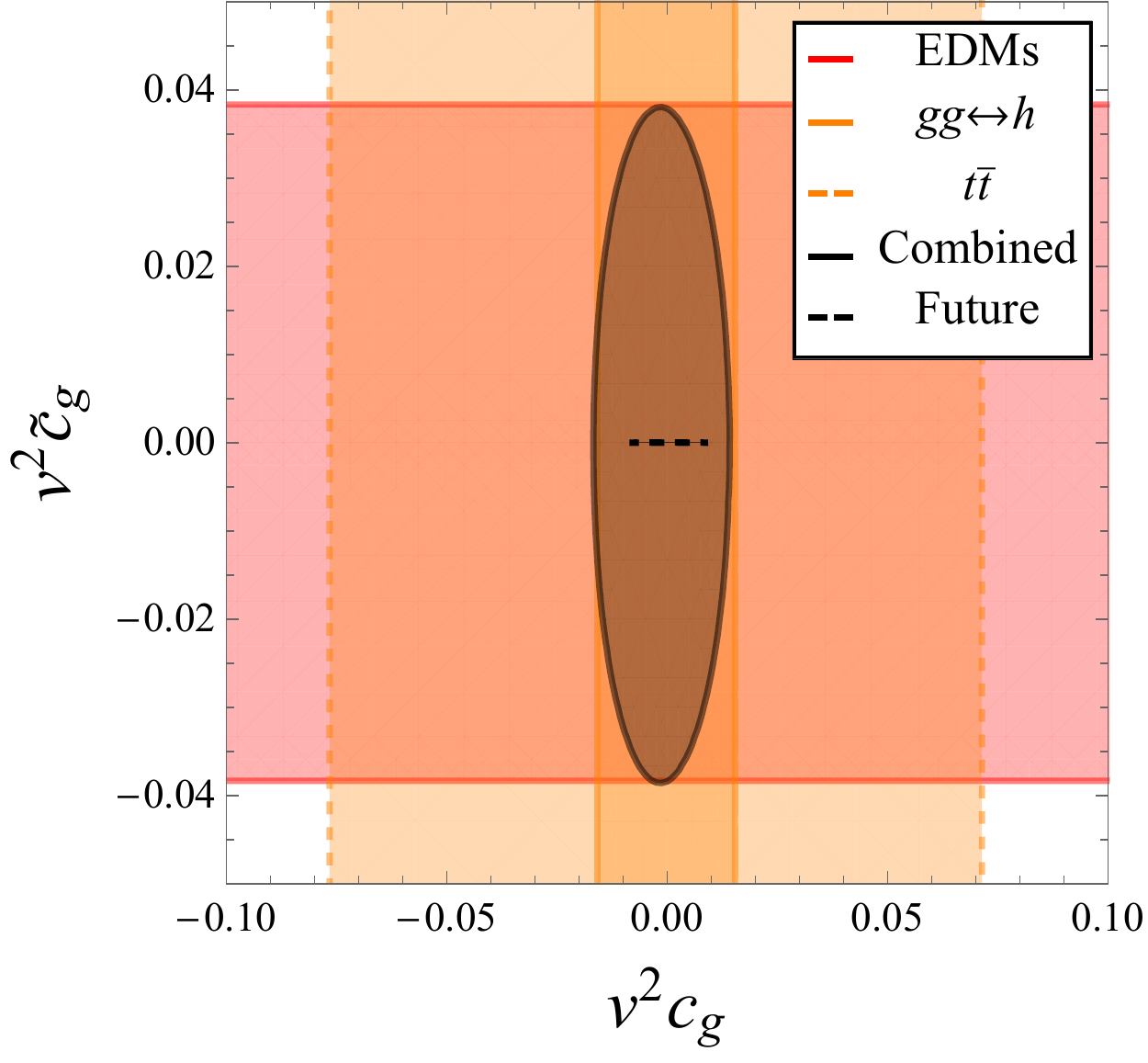}
\hspace{0.03\linewidth}
\includegraphics[width=0.3\linewidth]{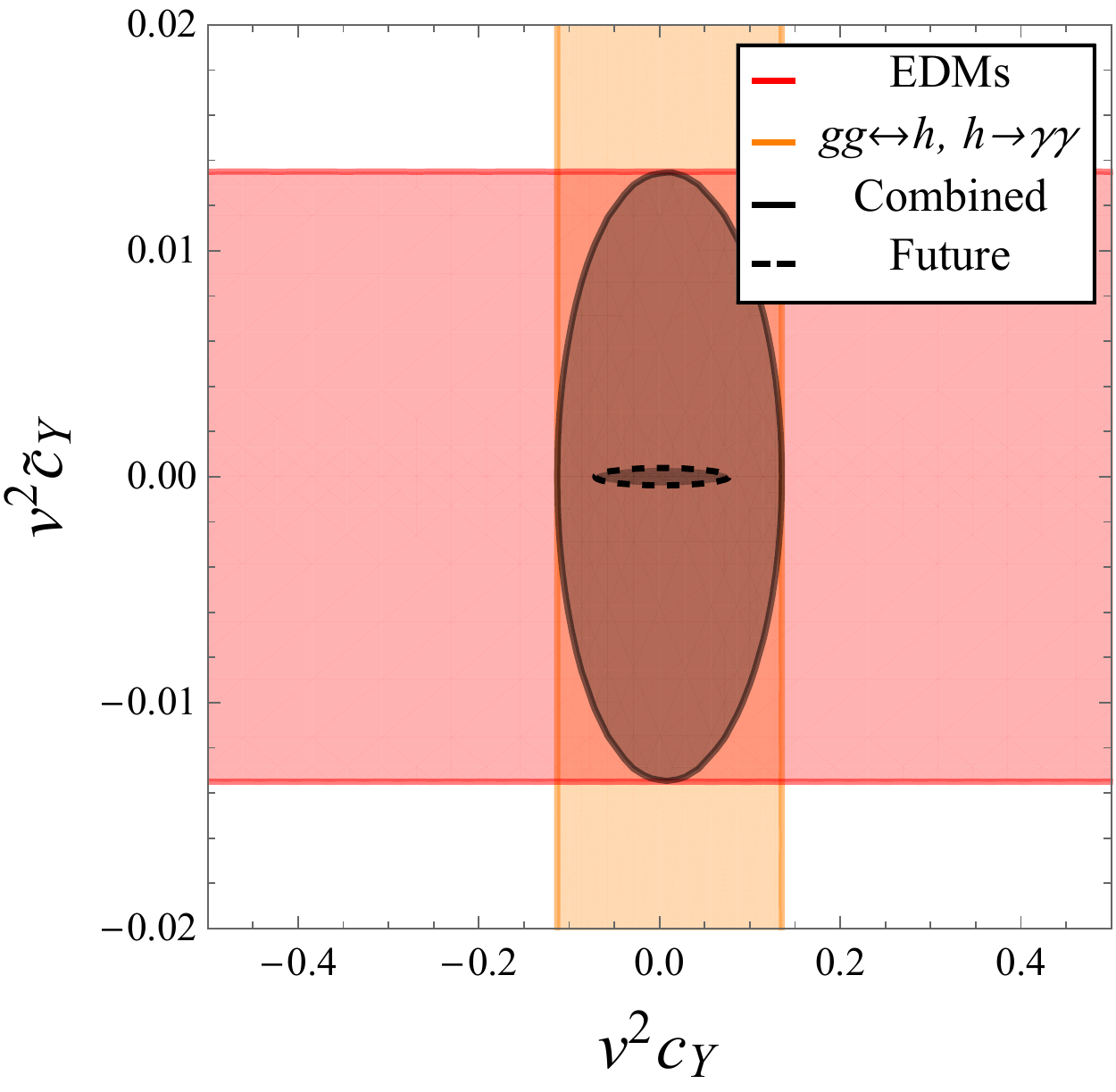}
\caption{90\% CL  allowed regions in the   $v^2 c_{Wb} - v^2  \tilde{c}_{Wb}$   (left panel) 
and $v^2 c_{g} - v^2  \tilde{c}_{g}$  (center panel) 
and $v^2 c_{Y} - v^2  \tilde{c}_{Y}$ planes (right panel).
with couplings evaluated at   $\Lambda = 1$~TeV.
Future EDM searches will probe $v^2 \tilde{c}_{Wb} \sim 2 \cdot  10^{-4}$, $v^2 \tilde{c}_{g} \sim 2\cdot 10^{-3}$, and $v^2 \tilde{c}_{Y} \sim  3 \cdot 10^{-4}$.
\label{fig:CWbgY}}
\end{figure*}

We first  focus on the case in which a single operator structure  dominates at the 
high scale,  keeping both real  ($c_\alpha$) and imaginary 
($\tilde{c}_\alpha$)  parts.  
In Figs.~\ref{fig:CgammaCWt} and  \ref{fig:CWbgY}  
we present the  90\% CL bounds on the  planes $v^2 c_{\alpha} - v^2  \tilde{c}_\alpha$, 
for  the five couplings of Eq.~(\ref{eq:Leff}).
We show the individual most constraining bounds  and the combined allowed region. 
For $\tilde{c}_\gamma,$ $\tilde{c}_{Wt}$, $\tilde{c}_{Wb}$, and $c_g$, 
our bounds are considerably stronger than the existing literature. For the remaining couplings, we agree with previous findings.
The following features emerge from the plots:
(i)  Indirect  probes  are currently  more constraining than  direct ones,   
with the exception of $c_{Wt}$, for which the bound from $W$ helicity fractions 
in $t \to W b$ competes with $b \to s \gamma$.
In particular, the bound on $c_g$ from Higgs production is a factor of 5 stronger than the direct bound from $t\bar t$.
(ii) EDMs, despite the conservative nature of the range-fit procedure, strongly constrain the CPV couplings, 
with the electron EDM dominating the bound on 
$\tilde{c}_\gamma$, 
$\tilde{c}_{Wt}$,  
$\tilde{c}_Y$, and $\tilde{c}_g$    
and the neutron EDM leading to the best bound on 
$\tilde{c}_{Wb}$. The neutron EDM could put a much stronger constraint on $\tilde c_g$ ($v^2 \tilde c_g < 2 \cdot 10^{-3}$) with better control of the hadronic matrix elements.
In particular EDMs lead to a   three (two) orders of magnitude 
improvement in the bounds on  $\tilde{c}_{\gamma}$  ($\tilde{c}_ {Wt}$), see Fig.~\ref{fig:CgammaCWt}, 
and a significant one (factor of 5) in $\tilde{c}_{Wb}$, see Fig.~\ref{fig:CWbgY}. The new bounds on $\tilde{c}_{\gamma}$ and $\tilde{c}_ {Wt}$ lie well below the prospected sensitivities of the LHC \cite{Fael:2013ira,Rontsch:2015una,Hioki:2015env} and envisioned \cite{Bouzas:2013jha,Rontsch:2015una} collider experiments.

In  Figs.~\ref{fig:CgammaCWt} and~\ref{fig:CWbgY}  
we also present projected combined 
bounds for the new physics couplings,  
based on  expected improvements in collider~\cite{CMS:2013xfa,ATL-PHYS-PUB-2014-016}, 
super-B factory~\cite{Bona:2007qt,Nishida:2011dh},  and EDM sensitivities~\cite{Kumar:2013qya} (one (two) order(s) of magnitude for the electron (neutron)). 

\textbf{Discussion:} 
\begin{figure}[t]
\centering
\includegraphics[width=0.8\linewidth]{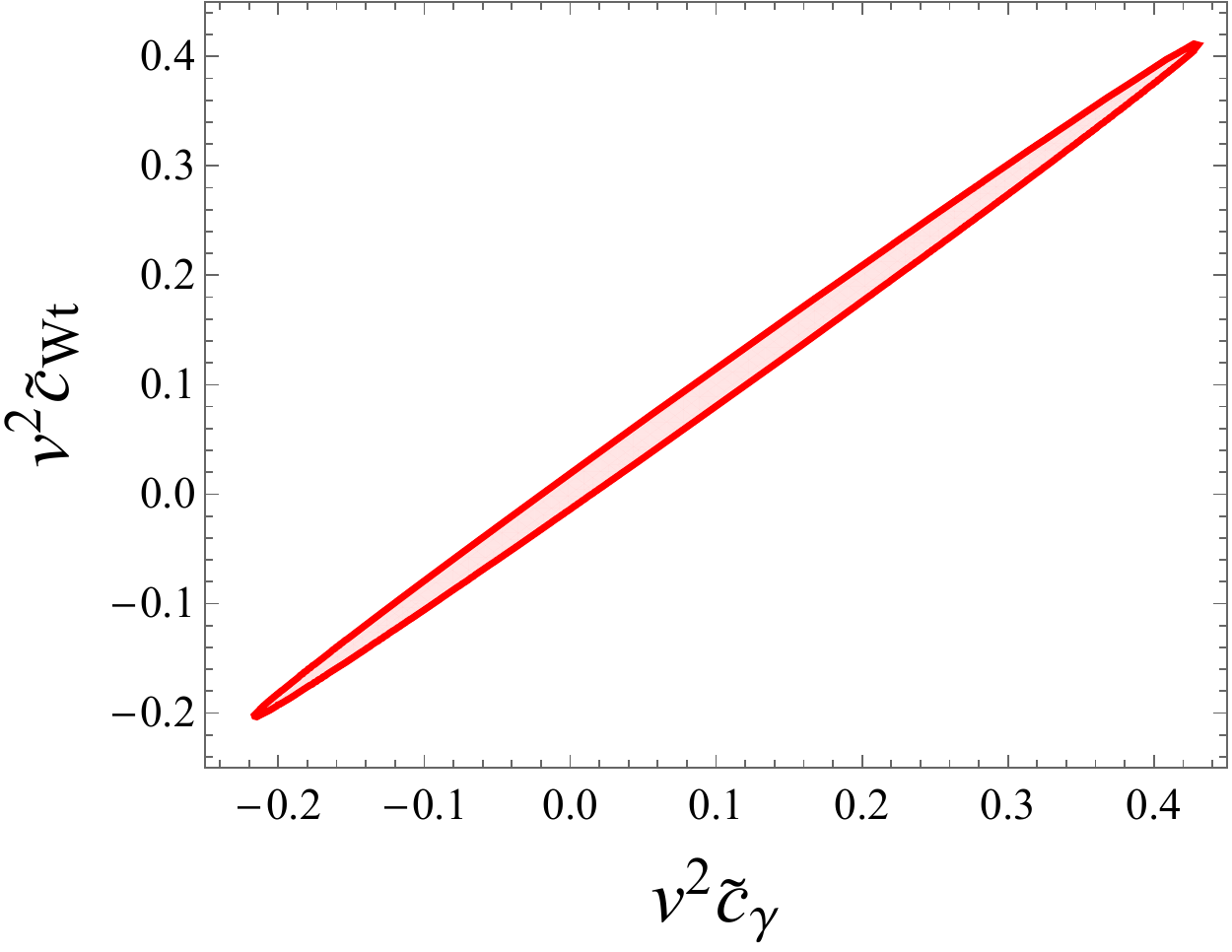}
\caption{90\% CL  allowed region in the   $v^2 \tilde{c}_{\gamma} - v^2  \tilde{c}_{Wt}$ plane 
in a global analysis.}
\label{correlations}
\end{figure}
The overarching message emerging from our single-operator analysis is that the 
CPV couplings are very tightly constrained, and  out of reach of direct  collider searches. 
If new physics   simultaneously generates  several operators at the scale $\Lambda$, 
not necessarily involving top and Higgs fields, our  results enforce strong correlation between 
the various couplings.  For example,   a large top EDM ($\tilde{c}_\gamma$) is compatible with 
non-observation of ThO EDM  if an electron EDM ($d_e$) is also generated at the  scale $\Lambda$, 
with the right size to  cancel the RG  effect from  $\tilde{c}_\gamma$, at the level of a few parts in a thousand.     
This puts  powerful constraints on the underlying dynamics,   
providing non-trivial  input  to model building~\cite{McKeen:2012av}.  
This point can also be illustrated by studying the case in which new physics generates 
all the couplings of Eq.~(\ref{eq:Leff}) at the matching scale $\Lambda$.
Performing a global analysis with  five  free CPV couplings $\tilde{c}_\alpha (\Lambda)$ 
(fixing the hadronic and nuclear matrix elements to  their central values) 
we find the bounds: 
 $-0.2 < v^2 \tilde c_\gamma < 0.4$, 
 $-0.02 < v^2 \tilde c_g < 0.04$, 
 $ -0.2 < v^2 \tilde c_{Wt} < 0.4$, 
 $-0.1 < v^2 \tilde c_{Wb} < 0.3$, 
  $-0.2 < v^2 \tilde c_{Y} < 0.5$. 
While  weaker than the single-operator EDM constraints,   in most cases these bounds  
are still stronger than individual flavor and collider bounds, 
and certain directions in parameter space remain very strongly constrained,  as shown in Fig. \ref{correlations}.
Even if one allows for additional CP violation in chirality-conserving top couplings, under broad assumptions such as MFV 
this strong constraint is not significantly affected.

\textbf{Conclusions:} 
In this letter we have highlighted   the impact of indirect  probes on chirality-flipping top-Higgs couplings, 
uncovering  the dramatic effect of neutron and atomic/molecular EDMs -- 
they improve the bounds on the top EDM  by  three  orders of magnitude ($|d_t| <  5 \cdot 10^{-20}$ e cm at $90\%$ C.L.).
Our results have implications for  baryogenesis mechanisms, collider searches, and  flavor physics.  
They motivate more sensitive EDM searches and improved lattice QCD and nuclear structure calculations of the effect of 
CPV operators in nucleons and nuclei.

\section*{Acknowledgements}
VC and EM  acknowledge support by the US DOE Office of Nuclear Physics and by the LDRD program at Los Alamos National Laboratory.
WD and JdV  acknowledge  support by the Dutch Organization for Scientific Research (NWO) 
through a RUBICON  and VENI grant, respectively.

\bibliographystyle{h-physrev3} 
\bibliography{bibliography}

\end{document}